# Maximum spin polarization in chromium dimer cations demonstrated by X-ray magnetic circular dichroism spectroscopy

Vicente Zamudio-Bayer,[a] Konstantin Hirsch,[a] Andreas Langenberg,[a] Markus Niemeyer,[b] Marlene Vogel,[a] Arkadiusz Ławicki,[a] Akira Terasaki,[c] J. Tobias Lau,*[a] and Bernd von Issendorff [d]

**Abstract:** X-ray magnetic circular dichroism spectroscopy has been used to characterize the electronic structure and magnetic moment of $Cr_2^+$. Our results indicate that the removal of a single electron from the $4s\sigma_g$ bonding orbital of $Cr_2$ drastically changes the preferred coupling of the 3d electronic spins. While the neutral molecule has a zero-spin ground state with a very short bond length, the molecular cation exhibits a ferromagnetically coupled ground state with the highest possible spin of S = 11/2, and almost twice the bond length of the neutral molecule. This spin configuration can be interpreted as a result of indirect exchange coupling between the 3d electrons of the two atoms that is mediated by the single 4s electron via strong intra-atomic 3d–4s exchange interaction. Our finding allows for an estimate of the relative energies of two states that are often discussed as ground state candidates, the ferromagnetically coupled $^{12}\Sigma$ and the low-spin $^2\Sigma$ state.

Even though homonuclear diatomic molecules are the smallest and simplest molecular entities, their electronic structure can already be highly complex to a degree that is demanding even for state-of-the-art electronic structure theory. A good example are diatomic molecules of the 3d transition elements[1-4] where the open 3d shell leads to a large number of possible electronic configurations. Among these molecules, $Cr_2$ and $Mn_2$ with their half-filled 3d shells represent extreme cases of electronic structure and bonding: $Mn_2$ is a weakly-bound molecule with a low-spin ground state and a large equilibrium distance[5-11] of 3.3-3.4 Å, whereas $Cr_2$ also is a low-spin molecule but with a sextuple bond and a short equilibrium distance[12-14] of 1.68 Å. Because of the particular challenge to correctly describe this rather weak sextuple bond, the ground state of $Cr_2$ is often used as a benchmark for electronic structure calculations[15-22]. While the neutral $Cr_2$ molecule by now is fairly well understood, there is a surprising lack of spectroscopic information on the $Cr_2^+$ molecular cation, whose ground state is still unknown. Only its binding energy has been determined experimentally[14,23], yielding values of 1.30 ± 0.06 eV to 1.43 ± 0.05 eV. As this is very close to the 1.443 ± 0.056 eV binding energy[13,14] of neutral $Cr_2$, one might expect similar electronic structures of these two dimers. Based on their collision-induced dissociation studies of $Cr_2^+$, Su et al. have indeed proposed a $^2\Sigma$ low-spin ground state[14], which would arise from the $(3d\sigma_g)^2(3d\pi_u)^4(3d\delta_g)^4(4s\sigma_g)^2$ $^1\Sigma$ configuration of $Cr_2$ by removal of one electron from the $4s\sigma_g$ orbital. This view has been supported in a theoretical study by Gutsev et al. who have also obtained such a $^2\Sigma$ ground state[4] with an equilibrium distance of 1.66 ± 0.01 Å. Other calculations, however, predict a $^{12}\Sigma$ high-spin ground state with 2.91–2.96 Å equilibrium distance[24,25] for $Cr_2^+$.

Here we show experimentally that $Cr_2^+$ has a high-spin ground state and thus differs strongly from $Cr_2$ in its electronic configuration. To this end we have employed $L_{2,3}$ (2p→3d) x-ray magnetic circular dichroism (XMCD) spectroscopy, which is a standard method to probe the magnetization of 3d transition metal bulk compounds or surface structures, but here is applied to free $Cr_2^+$ ions stored in a cryogenic linear quadrupole Paul trap within a strong magnetic field[26-31]. See supporting Information for details.

XMCD spectroscopy detects the difference in x-ray absorption cross sections for left- and right circularly polarized light on a magnetized sample, and in the case of 2p → 3d transitions probes the 3d electron magnetic moment. Our experimental results for $Cr_2^+$ are shown in Fig. 1a. The uppermost trace shows the average of the spectra measured for both polarizations, which is identical to the spectrum measured with linear polarization and without applied magnetic field, as presented in an earlier publication[32]. The middle trace shows the spectra for the two circular polarizations individually, measured in an applied magnetic field of 5 T, while the lower trace is the XMCD asymmetry or difference of these two. As discussed earlier[32], significant information can already be obtained from the linear absorption spectrum. For comparison, the calculated linear absorption spectrum of a chromium atom with the clear fingerprint pattern of a $3d^5$ $^6S$ configuration[33] is shown next to the experimental spectrum. The 2p absorption spectrum of $Cr_2^+$ is nearly identical to this atomic spectrum, which demonstrates that the 3d states are very similar to the atomic ones and obviously do not participate in covalent bonding[32]. This implies that the 3d orbitals of the two atoms must have very little overlap, which immediately allows us to estimate a lower bound of the $Cr_2^+$ bond length: since the atomic 3d radial electron distribution decreases to < 1 % for r > 1.35 Å, one can expect an equilibrium distance $r_e$ > 2.7 Å for the dimer cation. This finding already excludes the $^2\Sigma$ state with 1.66 ± 0.01 Å bond length predicted by Gutsev et al. as the ground state[4] of $Cr_2^+$, and seems to hint

[a] Dr. J.T. Lau, Dr. M. Vogel, Dr. A. Ławicki, Dr. K. Hirsch, Dr. A. Langenberg, Dr. V. Zamudio-Bayer
Institut für Methoden und Instrumentierung der Forschung mit Synchrotronstrahlung
Helmholtz-Zentrum Berlin für Materialien und Energie GmbH
Albert-Einstein-Straße 15, 12489 Berlin (Germany)
E-mail: tobias.lau@helmholtz-berlin.de
[b] MSc. M. Niemeyer
Institut für Optik und Atomare Physik
Technische Universität Berlin
Hardenbergstraße 36, 10623 Berlin (Germany)
[c] Prof. Dr. A. Terasaki
Department of Chemistry
Kyushu University
6-10-1 Hakozaki, Higashi-ku, Fukuoka 812-8581 (Japan)
[d] Prof. Dr. B. von Issendorff
Physikalisches Institut
Universität Freiburg
Stefan-Meier-Straße 21, 79104 Freiburg (Germany)

Supporting information for this article is given via a link at the end of the document.

at the high spin state that is usually obtained as the lowest state at long bond lengths.

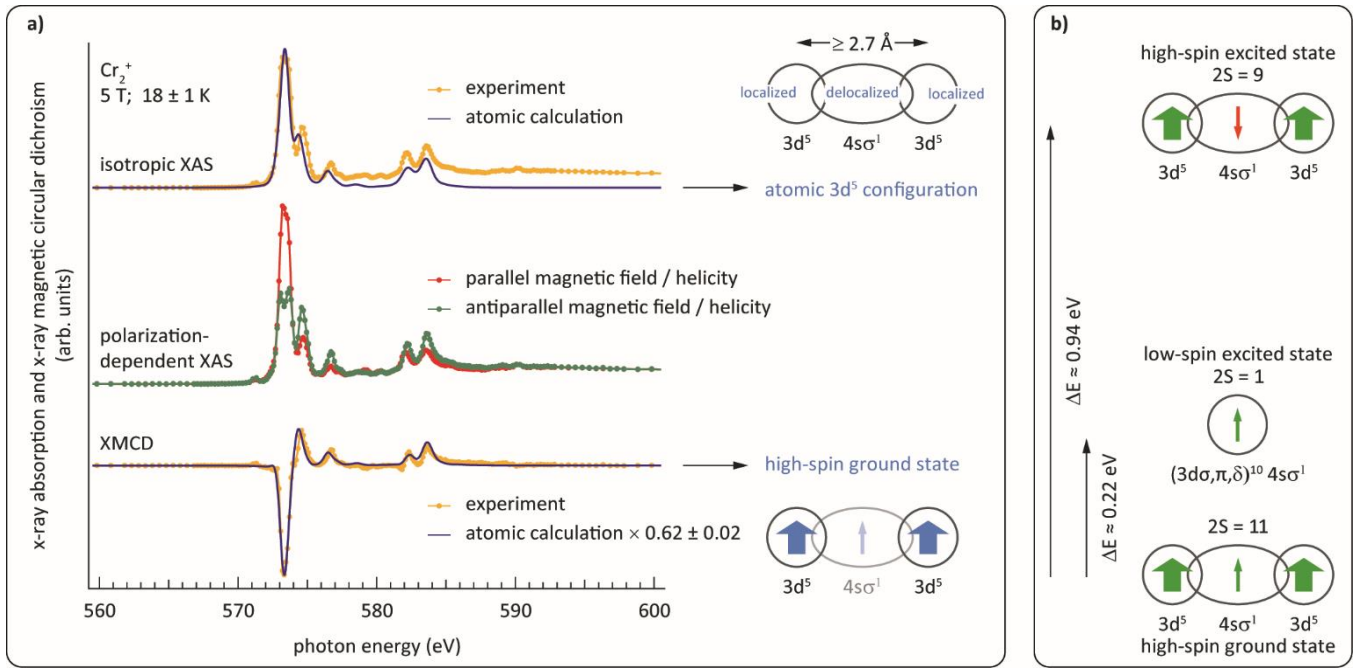

**Figure 1.** a) The close match between the calculated atomic fingerprint of the Cr $3d^5$ configuration (solid line) and the experimental isotropic 2p x-ray absorption spectrum of $Cr_2^+$ (data points) reveals atomic localization of the 3d orbitals and indicates that bonding is mediated by the single 4s electron [32]. The $Cr_2^+$ 2p x-ray magnetic circular dichroism asymmetry agrees well with a scaled atomic calculation from which a high-spin configuration with a net 3d magnetization of 6.2 ± 0.2 $\mu_B$ can be deduced. b) Relative energies of the $Cr_2^+$ low-spin (S = 1/2) and high spin (S = 9/2; $S_d$ = 10/2) excited states with respect to the S = 11/2 ground state. See text and Fig. 2 for details and relative energies provided.

The experimental proof of such a high spin configuration is delivered by the XMCD data. In the middle and lower trace of Fig. 1a one can see a substantial magnetic dichroism, which indicates a rather strong magnetic moment of the chromium dimer cation, and therefore a high spin state. This can even be quantified although the spin sum rule of x-ray magnetic circular dichroism[34] cannot be applied to $3d^n$ configurations with n ≤ 5, such as present in chromium[35,36]. Because of the local atomic $3d^5$ $^6S$ configuration in $Cr_2^+$ one can compare the experimental XMCD spectrum to a calculated one of the chromium atom. This should be a good approximation because of the weak perturbation of the 3d states observed in the x-ray absorption spectrum. Close to perfect agreement between experimental and calculated spectra is obtained with a fit that gives an alignment of (62 ± 2) % of the atomic 3d magnetic moment. Furthermore the orbital angular momentum sum rule of x-ray magnetic circular dichroism[37] yields $\mu_l$ = 0.05 ± 0.22 $\mu_B$, i.e., zero orbital angular momentum within the error bars. This means that we are observing a Σ state with a 3d magnetization of (0.62 ± 0.02)· 10 $\mu_B$ = 6.2 ± 0.2 $\mu_B$. In case of a $^{12}Σ$ state of $Cr_2^+$, where the spins of all 3d electrons and the single 4sσ electron are aligned in parallel to each other, such a magnetization of the 3d moments in a 5 T magnetic field would be obtained from the Brillouin function for an electronic temperature of 18 ± 1 K. This is exactly the temperature expected for the $Cr_2^+$ ion, as the experiment was performed at an ion trap temperature of 13 ± 1 K, and a radio frequency heating of the stored ions by 4–8 K is unavoidable under our experimental conditions[28,30,31]. The next lower spin states of the chromium dimer cation are two $^{10}Σ$ states, which differ by the relative alignment of the contributing spins. As can be expected and is also seen in the atom-like linear x-ray absorption spectrum, the 3d electrons on each atom form S = 5/2 states, which, together with the single spin of the 4sσ electron, couple to the total spin of $Cr_2^+$. The resulting states can be described by assuming that the two S = 5/2 spins couple to an overall 3d spin $S_d$, which then couples with the single 4sσ electron spin (this is not exactly true, but should be a good approximation). A total spin of S = 9/2 can thus be produced by coupling of the two S = 5/2 spins to a spin of $S_d$ = 10/2, with additional antiparallel coupling of the single 4sσ spin; or by coupling to an $S_d$ = 8/2 state, with parallel coupling of the 4sσ spin. Because of its lower 3d magnetic moment, the latter $^{10}Σ$ ($S_d$ = 8/2) state would exhibit a 3d magnetization of 6.2 ± 0.2 $\mu_B$ in a 5 T field only at a temperature of 10 ± 1 K and can therefore immediately be ruled out, as can all lower ($S_d$ ≤ 8/2) spin states. The former $^{10}Σ$ ($S_d$ =

10/2) state would reach the observed magnetization at an ion temperature of 15 ± 1 K, which is close to impossible to achieve in a 13 ± 1 K ion trap under our experimental conditions. But most importantly, as will be discussed in the following, this $^{10}\Sigma$ ($S_d$ = 10/2) state should be almost 1 eV higher in energy than the $^{12}\Sigma$ state, and hence can safely be ruled out. Our XMCD experiment therefore directly shows that the $^{12}\Sigma$ state is the true ground state of $Cr_2^+$.

The question arises as to the cause of this ferromagnetic coupling of the 3d electrons at a bond length where the 3d orbitals practically do not overlap. As has been discussed in detail by Bauschlicher[38] for the $(4s\sigma_g)^2(4s\sigma_u)^1(3d^5)(3d^5)$ $^{12}\Sigma$ ground state of $Mn_2^+$, this can be viewed as the result of indirect exchange coupling mediated by the shared $4s\sigma$ electron. The intra-atomic exchange interaction between the 3d electrons and the 4s electron is exceptionally high for chromium, and the second highest among the 3d transition elements: the energy difference[39] between the $^7S$ atomic ground state and the $^5S$ excited state, which differ by a spin flip of the 4s electron, is 0.94 eV. If the 4s electron occupies a $4s\sigma$ orbital and thus is equally shared between the two atoms, the energy difference between parallel and antiparallel alignment of this electron and the 3d electrons on a single atom should then be about 0.47 eV. The $^{10}\Sigma$ ($S_d$=10/2) state mentioned above should therefore be 0.94 eV higher in energy than the ground state, as here a change from parallel to antiparallel alignment occurs on both atoms.

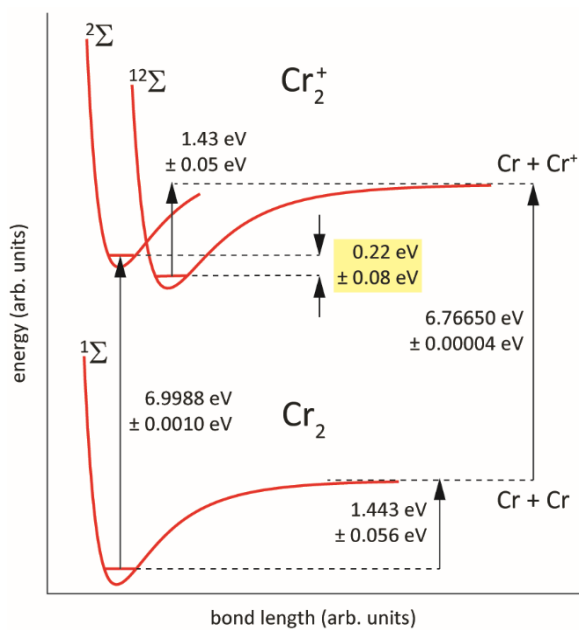

**Figure 2.** Schematic potential energy curves in a Born-Haber cycle to determine the relative energy of the $Cr_2^+$ low-spin $^2\Sigma$ ($S_d$ = 0) and high-spin $^{12}\Sigma$ ($S_d$ = 10/2) states. The high-spin state is more stable than the low-spin state by 0.22 ± 0.08 eV. This cycle also gives 6.79 ± 0.05 eV as the adiabatic ionization potential of $Cr_2$. Experimental ionization potentials and bond dissociation energies are taken from Refs. [13,23,40,41]. This is a strongly simplified scheme. For a more realistic representation of all low lying states of $Cr_2$ see, e.g., Anderson.[42] The potential energy curves of $Cr_2^+$ will be similarly complex.

This identification of a maximum spin ground state of $Cr_2^+$ allows us to reassess a thermodynamic cycle[14,41] for an estimate of the relative energy between the potential energy minima of the $^{12}\Sigma$ ground state and the low-lying $^2\Sigma$ excited state from available literature values of ionization potentials[40,41] and bond energies[13,14,23]. These values can be combined in a Born-Haber cycle, as shown in Fig. 2: Photoionization of neutral $^1\Sigma$ $Cr_2$ produces $^2\Sigma$ $Cr_2^+$, which has a similar bond length and can be reached without a spin flip in a vertical transition[41]. This $^2\Sigma$ state of $Cr_2^+$ is now identified as an excited state by our results. The binding energy of $Cr_2^+$, however, has been measured for its ground state, that is for the $^{12}\Sigma$ state[14,23]. In combination with the ionization potential of the Cr atom[40] and the experimentally determined binding energy[13,14] of $Cr_2$, we can thus give the energy difference between the $^{12}\Sigma$ and $^2\Sigma$ states of $Cr_2^+$ to 0.22 ± 0.08 eV. This value is much smaller than the 1.33 eV obtained in the recent "gold standard" calculations[25] by Yamada et al., illustrating the difficulty to correctly describe the chromium dimer cation even at the highest levels of theory. As for $Cr_2$, where the corresponding $^1\Sigma$ and $^{11}\Sigma$ states are reversed in energy, the existence of a low-lying excited state of $Cr_2^+$ leads to complications in theory and experiment[12-22,42].

We would like to conclude this discussion by remarking that, similar to the case of $Mn_2^+$,[5,6,38,43,44] bonding and spin coupling in $Cr_2^+$ can be viewed in the Zener solid-state double-exchange model[45,46]. In both molecular cations, the unpaired spin of one electron in a singly occupied 4s-derived molecular orbital mediates ferromagnetic coupling of localized 3d electrons in high-spin states via strong intra-atomic spin correlation.

In summary, the $^{12}\Sigma$ ground state of the $Cr_2^+$ diatomic molecular cation has been identified experimentally by x-ray magnetic circular dichroism spectroscopy. Based on a reassessment of the thermodynamic cycle, this $^{12}\Sigma$ state should be about 0.22 ± 0.08 eV below the lowest $^2\Sigma$ state, which is often assumed to be the ground state. In $Cr_2$ and $Cr_2^+$ two extreme cases of the electronic configuration are realized, with complete spin pairing of bonding 3d electrons in singlet $Cr_2$, and fully unpaired spins of localized 3d electrons in dodecaplet $Cr_2^+$. Mixed valence $Cr_2^+$ is the simplest case of s–d exchange coupling of localized 3d high spin states by a single 4s electron and can thus be regarded as a model system for indirect exchange. Bonding in $Cr_2^+$ can be understood in analogy to $Mn_2^+$, even though the corresponding neutral species are very different[38]. Our result shows that, under the special circumstance of large bond lengths, molecules of 3d transition elements with half-filled 3d shells might behave like rare earth elements with their strongly localized 4f electrons.

## Acknowledgements

Beam time for this project was granted at BESSY II beamlines UE52-SGM and UE52-PGM, operated by Helmholtz-Zentrum Berlin. The superconducting solenoid was provided by Toyota Technological Institute. AT acknowledges financial support by Genesis Research Institute, Inc. BvI acknowledges travel support by Helmholtz-Zentrum Berlin. This project was partially funded by the German Federal Ministry of Education and Research (BMBF) through grant BMBF-05K13Vf2.

**Keywords:** Exchange interactions • Magnetic properties • Chromium • X-ray absorption spectroscopy • Ion trap

**Entry for the Table of Contents**

## COMMUNICATION

*Vicente Zamudio-Bayer, Konstantin Hirsch, Andreas Langenberg, Markus Niemeyer, Marlene Vogel, Arkadiusz Ławicki, Akira Terasaki, J. Tobias Lau,\* and Bernd von Issendorff*

*Page No. – Page No.*

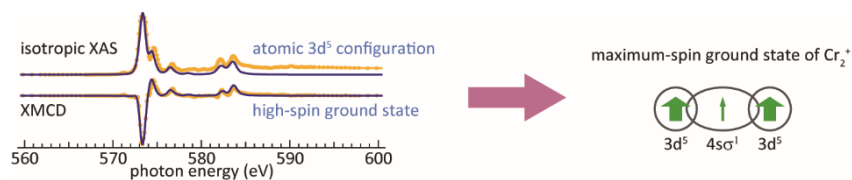

**Maximum spin polarization in chromium dimer cations demonstrated by X-ray magnetic circular dichroism spectroscopy**

The removal of a single electron from the $4s\sigma_g$ bonding orbital of $Cr_2$ fully localizes all 3d electrons and drastically changes the preferred coupling of their spins. The molecular cation exhibits a ferromagnetically coupled ground state with the highest possible spin of $S = 11/2$, and almost twice the bond length of the neutral molecule. This spin configuration can be interpreted as a result of indirect exchange coupling.